\documentclass[12pt]{article}
\usepackage{amsfonts}
\usepackage{amsmath}
\usepackage{amssymb}

\def\be{\begin{equation}}
\def\ee{\end{equation}}
\def\bea{\begin{eqnarray}}
\def\eea{\end{eqnarray}}
\newcommand{\nn}{\nonumber}
\newcommand{\kP}{$\kappa$-Poincar\'e}
\newcommand{\kM}{$\kappa$-Minkowski}
\newcommand{\al}{\alpha}
\newcommand{\bt}{\beta}
\newcommand{\x}{{\bf{\hat{x}}}}
\newcommand{\g}{\mathfrak{g}}

\newcommand\opname[1]{\mathop{\mathrm{#1}}\nolimits}
\newcommand{\tr}{\opname{tr}}

\begin{document}

\title{\Large {\bf $\kappa$-Minkowski representations on Hilbert spaces}}
\maketitle

\begin{center}
\author{\bf Alessandra Agostini\footnote{Supported by Quantum
Spaces-Noncommutative Geometry EU Network}}
\end{center}

\begin{center}
{School of Mathematics, Cardiff University, \\
Senghennydd Road, Cardiff CF24 4AG, Wales, United Kingdom}
\end{center}

\abstract{The algebra of functions on \kM\ noncommutative spacetime
is studied as algebra of operators on Hilbert spaces. The
representations of this algebra are constructed and classified. This
new approach leads to a natural construction of integration in
$\kappa$-Minkowski spacetime in terms of the usual trace of
operators.}

\maketitle


\section{\label{sec:1}Introduction}

Some approaches to Quantum Gravity \cite{DFR,garay} are based on the
idea that at Planck-length distances geometry can have a form quite
different from the one we are familiar with at large scales. In
particular, Doplicher et al.~\cite{DFR} explored the possibility
that Quantum Gravity corrections can be described algebraically by
replacing the traditional (Minkowski) spacetime coordinates
$x_{\mu}$ with \emph{Hermitian operators} $\x_{\mu}$
($\mu,\nu=0,1,2,3$) which satisfy nontrivial commutation relations
$$ [\x_{\mu},\x_{\nu}]=i\theta_{\mu\nu}(\x).
$$
 A noncommutative spacetime of this type embodies an impossibility to fully know the
short distance structure of spacetime, in the same way that in the
phase space of the ordinary \emph{Quantum Mechanics} there is a
limit on the localization of a particle.

However, the idea of the spacetime noncommutativity has a more
ancient origin which goes back to Heisenberg and was published for
the fist time by Snyder~\cite{Snyder1,Snyder2}. The motivation at
that time was the hope that noncommutativity among coordinates could
improve the singularity of quantum field theory at short-distances
\cite{Jackiw:2002tw}.

Noncommutative geometry emerges also at the level of effective
theories, for example in the description of Strings in the presence
of external fields~\cite{SW,douglasnovikov}, or in the description
of electronic systems in the presence of external
magnetic-background \cite{Jackiw:2001dj}.

There is a wide literature on the simplest ``canonical"
noncommutativity characterized by a constant value of the
commutators $$ [\x_{\mu},\x_{\nu}]=i\theta_{\mu\nu}, $$ where
$\theta_{\mu\nu}$ is a matrix of dimensionful parameters. This
noncommutative spacetime arises in the description of string
M-theory in presence of external fields~\cite{CDS}.


In this paper we consider another much studied noncommutative
spacetime, called \kM\ spacetime, characterized by the commutation
relations $$ [\x_0,\x_j]=i\lambda\x_j\;\;\;[\x_j,\x_k]=0,
\;\;\;j=1,2,3 $$ where $\lambda\in \mathbb{R}_{\setminus 0}$
represents the noncommutativity parameter\footnote{Historically, the
noncommutative parameter $\kappa=\lambda^{-1}$ was introduced. This
explains the origin of the name ``\kM".}. To be simple we shall
consider $\lambda>0$ in this paper. This type of noncommutativity,
introduced in~\cite{MajidRuegg}, is an example of Lie-algebra
noncommutativity where the commutation relations among spacetime
coordinates exhibit a linear dependence on the spacetime coordinates
themselves $$ [\x_{\mu},\x_{\nu}]=i \zeta^\rho_{\mu\nu}\x_\rho, $$
with coordinate-independent $\zeta_{\mu\nu}^\rho$. This algebra was
proposed in the framework of the Planck scale
Physics~\cite{Maj88MajPhDmaj00} as a natural candidate for a
quantized spacetime in the zero-curvature limit.

Recently, \kM\ gained remarkable attention due to the fact that it
provides an example of noncommutative spacetime in which Lorentz
symmetries are preserved as deformed (quantum) symmetries
\cite{aad03, Dimitrijevic:2003pn, Ballesteros:2003ag}. In particular,
Majid\&Ruegg~\cite{MajidRuegg} characterized the symmetries of
\kM\ with the so called \kP\ algebra \cite{lukieAnnPhys} already
known in the Quantum Group literature as significant example of
deformation of the Poincar\'e algebra.

The analysis of the physical implications of the deformed \kP\
algebra has led to interesting hypotheses about the possibility
that in \kM\ particles are submitted to modified dispersion
relations~\cite{AmelinoLukNowik}
which agree with the postulates of Doubly Special Relativity
theories \cite{dsr1,Agostini:2004cu}, recent relativistic theories
with both an observer-independent velocity scale and an
observer-independent length scale (possibly given by the Planck
length).

Over the past few years there has been a growing interest in the
construction of field theories on \kM\ (see for example
\cite{Dimitrijevic:2003pn,Agostini:2003dc}). The approach to this
study has essentially been based on the introduction of a
deformation of the product among the coordinate functions. This
deformed product (called \emph{star-product}) replaces the
commutative product and makes it possible to map a noncommutative
theory into a theory with commutative functions multiplied through
the deformed product~\cite{Madore:2000en}. In this way an analysis
of field theories in noncommutative spacetime results to be quite
similar to the one adopted in the ordinary commutative spaces.

However, several technical difficulties are encountered in this
construction, and the results obtained so far are still partial,
especially in comparison with the result obtained for the
canonical noncommutative spacetime \cite{Minwalla}.

In this paper we propose a new approach to the study of \kM\
spacetime based on the analysis of \kM\ algebra as algebra of
operators represented on Hilbert space. Besides the interest this
analysis provides by itself, it might also prove useful in
understanding the source of the technical problems encountered so
far in the construction of a field theory on \kM.

We write the fields in \kM\ as Fourier expansion in \emph{plane
waves}, as we usually do in commutative field theory, with the
difference that in this case the plane waves are functions of
noncommutative coordinates. These waves are shown to be the
elements of a unitary Lie group corresponding to the \kM\ Lie
algebra. In this way the problem of the representations of the
\kM\ fields reduces to the problem of the representation of a Lie
group.

We show that it is possible to obtain a Schr\"odinger representation
of this group on $L^2(\mathbb{R})$ by introducing Jordan-Schwinger
(JS) maps~\cite{selene} between the Quantum Mechanics operators of
position and momentum and the generators of \kM\ algebra.

The Schr\"odinger representation is the simplest one but is not the
only one possible. We study the problem of the existence and the
classification of the other representations of \kM\ group.

Since \kM\ group is a semidirect-product group, by using the
technique of induced representations, we are able to classify all
unitary irreducible representations (UIRs) of it. We discover that
they are all unitarity equivalent to two classes of Schr\"odinger
representations.

As the \kM\ fields can be represented as operators on Hilbert
space they form a $C^*$-algebra and we show that, under some
hypotheses, this is the $C^*$-algebra of compact operators on
$L^2(R,d\mu)$, with a particular choice of nontrivial measure
$d\mu$.

Moreover the knowledge of representations of \kM\ algebra leads to
a natural construction of integration in \kM\ as trace of
operators. We show that in this way we recover a proposal of
cyclic integration recently obtained in literature with different
approaches.

The paper is organized as follows. In Sec.~\ref{sec:2} we introduce
the algebra of \kM\ spacetime.
In Sec.~\ref{sec:3} we construct the Lie group corresponding to the
\kM\ algebra. Section~\ref{sec:4} is devoted to the representation
theory. In the first part of the section we use the JS maps to
obtain the Schr\"odinger representation (on $L^2(R)$) of the \kM\
group. In the second part of the section the technique of the
induced representations allows us to  prove that all IURs of the
\kM\ group are unitarily equivalent to the Schr\"odinger one. In
Sec.~\ref{Cstaralgebras} we discuss the properties of the \kM\
$C^*$-algebra. Finally, in Sec.~\ref{cyclic} we construct an
integral in \kM\ using the trace of operators.


\section{\label{sec:2}\kM\ Spacetime}

Let us consider the four-dimensional Minkowski spacetime
$\mathbb{\mathbb{R}}^{(3,1)}$. It can be viewed as a vector space
with basis given by the \emph{real} coordinates
$x_\mu$\footnote{In this paper Greek indices take values from zero
to three while Latin indices take values from one to three. The
Minkowski metric is taken to be of signature (+ - - - ).
}. This vector space can be made into a (Abelian) Lie algebra
$\mathfrak{g}_0$
 by introducing a trivial bilinear operation (Lie bracket) $[,]:\,\mathfrak{g}_0\times \mathfrak{g}_0\to
 \mathfrak{g}_0$
 \be
 [x_\mu,x_\nu]=0 \label{mink}.
 \ee
Now let us consider the flat vector space
$\mathbb{\mathbb{R}}^{(3,1)}$ with basis $\x_\mu$ and replace the
trivial Lie bracket (\ref{mink}) with the following one
  \bea [\x_0,\x_j]=i\lambda\x_j &&[\x_j,\x_k]=0,\;\;\;\;\;j,k=1,2,3. \label{kmink}\eea The flat vector space
$\mathbb{R}^{(3,1)}$ endowed with this Lie bracket is called \kM\
spacetime. The \kM\ coordinates $\x_\mu$ with relations
(\ref{kmink}) generate the \kM\ Lie algebra $\g_\lambda$. In the
limit $\lambda\to 0$ the commutative Minkowski algebra is
recovered \be \lim_{\lambda\to 0}\g_\lambda=\g_0\ee

As we have mentioned in the Introduction, we are interested in the
extension of the notion of classical fields\footnote{Even though
we are dealing with classical fields a sort of quantization
naturally arises from the noncommutativity of the spacetime
coordinates, but this quantization has not the meaning of the
second-quantization in particle Physics.} on \kM\ noncommutative
spacetime.

It is well known that a classical field on the commutative Minkowski
spacetime $\mathbb{R}^{(3,1)}$ can be represented as a Fourier
expansion in plane waves \be f(x)=\int d^3k dw\, \tilde{f}(w,k)
\,e^{iwx_0-ik_jx_j},\label{field}\ee where $\tilde{f}(w,k)$ is the
standard Fourier transform of the function $f(x)$, and the function
$e^{iwx_0-ik_jx_j}$ has the meaning of a plane wave.

Inspired by the commutative case we would like to write the
noncommutative fields in \kM\ as a Fourier expansion in
noncommutative exponential functions. But because of the
noncommutativity of the coordinates $\x_\mu$ there is an ambiguity
in generalizing the exponential function $e^{i(wx_0-k_jx_j)}$ to
the noncommutative case. One possibility is the time-to-the-right
exponential\be \x_\mu\to e^{-ik_j\x_j}e^{iw\x_0}\ee but the
alternative choice $e^{ik_\mu\x^\mu}$ and many others are also
possible. However, since all these maps are connected by a change
of variables $k_\mu$ (see \cite{aad03}) the particular choice of
map is not relevant for our study. In this paper we prefer to base
our formalism on the time-to-the right notation only because it
involves some advantages in the calculations.

Using the time-to-the right exponential the \kM\ generalization of
(\ref{field}) takes the form \be F(\x)\equiv\mathcal{W}(f(x))=\int
d^3k\,dw\; \tilde{f}(w,k)\, e^{-ik_j\x_j}e^{iw\x_0}
\label{KWeylmap}\ee where $\tilde{f}(w,k)$ is the classical
Fourier transform of the commutative function $f(x)$. The
commutative case suggests the interpretation of the noncommutative
functions $e^{-ik_j\x_j}e^{iw\x_0}$ as noncommutative plane waves.

This ``quantization" is conceptually similar to the Quantum
Mechanics quantization (see~\cite{AgarwalWolf}) for a detailed
description). The basic observables of Quantum Mechanics
($Q_j,P_j$) satisfy the Heisenberg algebra (or canonical)
commutation relations \be
[Q_i,P_j]=i\hbar\;\;\;[Q_i,Q_j]=[P_i,P_j]=0,\;\;\;i,j=1,2,...n
\label{heisenbergalgera}\ee
and the Weyl quantization of the Phase Space is given by the
following map \be F(Q,P)\equiv W(f)=\int d^n\al d^n\bt\,
\tilde{f}(\al,\bt)\, e^{i\al_j Q_j+i\bt_j P_j} \label{compatti} \ee
which  has exactly the same structure of the map in Eq.
(\ref{KWeylmap}). The operators $W(f)$ are called Weyl operators. In
analogy with this name we call the operators $\mathcal{W}(f)$
$\kappa$-Weyl operators.

The structure of the algebra of Weyl operators (\ref{compatti})
has been largely investigated in literature. It is a $C^*$-algebra
of the compact operators on $L^2(R)$, and its representations on
Hilbert space are very well-known.

The algebra of $\kappa$-Weyl operators (\ref{KWeylmap}) instead is
quite new and there are no studies about its representations on
Hilbert spaces. The purpose of the present paper is to investigate
this problem. Since the $\kappa$-Weyl operators are Fourier
expansions in the elements $e^{-ik_j\x_j}e^{iw\x_0}$ we are
essentially interested in the representations of these elements. As
we will show in the next Section they are the elements of a unitary
Lie group associated with the \kM\ Lie algebra $\g_\lambda$. Thus we
are essentially interested in representing this group.

\section{\label{sec:3}\kM\ Lie Group}
In this section the Lie groups of the \kM\ Lie algebra $\g_\lambda$
are constructed. The technique we use is based on the
Baker-Cambpell-Hausdorff (BCH) formula and represents a
generalization of the technique employed in the construction of the
Heisenberg group (\ref{heisenbergalgera}). We start this section by
reviewing the construction of the Heisenberg group and then we
generalize the procedure to the case of \kM.

\subsection{The Heisenberg Group}
The Heisenberg algebra (\ref{heisenbergalgera})
can be reformulated as a Lie algebra by introducing a further
basis vector $C$ such that \be
[Q_j,Q_k]=[P_j,P_k]=[Q_j,C]=[P_j,C]=0,\;\;\;[Q_j,P_k]=i\delta_{jk}C
\label{LieHeisenberg} \ee the relation to the previous commutation
relations (\ref{heisenbergalgera}) being that they correspond to
the case $C$ acts by $\hbar$. Introducing the generators
$T(\al,\bt,\gamma)=\sum_j(\al_j Q_j+\bt_j P_j+\gamma C)$,
Eq.~(\ref{LieHeisenberg}) reads
$$
[T(\al,\bt,\gamma),T(\al',\bt'\gamma')]=T(0,0,i(\al\bt'-\al'\bt)$$ A
Lie group associated with the Heisenberg algebra can be obtained
using the BCH formula (see~\cite{Folland})
$$
e^{T(\al,\bt,\gamma)}\,e^{T(\al',\bt'\gamma')}=e^{T(\al,\bt\gamma)+T(\al',\bt'\gamma')+\frac{1}{2}T(0,0,i(\al\bt'-\al'\bt))}=e^{T(\al+\al',\bt+\bt'+\gamma+\gamma'+\frac{i}{2}(\al\bt'-\al'\bt))}
$$
The identification of the element $\al,\bt,\gamma\in
\mathbb{R}^{2n+1}$ with the element $e^{T(\al,\bt,\gamma)}$ makes
$\mathbb{R}^{2n+1}$ into a group, called Heisenberg group, with
composition law \be (\al,\bt,\gamma)(\al',\bt',\gamma')=(
\al+\al',\bt+\bt',\gamma+\gamma'+\frac{i}{2}(\al\bt'-\al'\bt))\label{HeisGroup}\ee

We notice that the elements $e^{i\sum_j\al_j Q_j+\bt_j P_j}$ which
appear in the Weyl-quantization map (\ref{compatti}) are elements
of a unitary Heisenberg group with $\gamma=0$.

\subsection{The \kM\ Group}
In Ref. \cite{Agostini:2002de} we have shown that the construction
of the Heisenberg group (\ref{HeisGroup}) can be generalized to any
Lie-algebra. The only difference is that the composition law is no
longer Abelian.

Here we show this result in the case of \kM\ using the same
procedure adopted in the Heisenberg case above. It is sufficient to
work in $1\!+\!1$-dimensions, the result being straightforwardly
extendable to any dimension.

Let us write the commutation relations of $1\!+\!1$-dimensional \kM\
algebra in the form $$ [(\al \x_0+\bt \x),(\al' \x_0+\bt'
\x)]=i\lambda (\al\bt'-\al'\bt)\x $$ By introducing the generators
$T=\al \x_0+\bt \x$ we see that $$
[T(\al,\bt),T(\al',\bt')]=i\lambda T(0,\al\bt'-\al'\bt). $$ As in
the case of the Heisenberg algebra, a Lie group associated with the
\kM\ algebra can be obtained by the BCH formula
$$
e^{T}\,e^{T'}=e^{T+T'+\frac{1}{2}[T,T']+\frac{1}{12}[T,[T,T']+[[T,T'],T']]+..},$$
where $T=T(\al,\bt)$ and $T'=T(\al',\bt')$. The series in the
exponential function at the right side has been
computed~\cite{Kosinski:1999dw} and the following relation has been
obtained \be e^{T(\al,\bt)}e^{T(\al',\bt')}=
e^{T(\al+\al',\,\frac{\bt\phi(\al)+e^{i\lambda
\al}\bt'\phi(\al')}{\phi(\al+\al')})}, \label{kappagroup}\ee  where
$\phi(\al)=i(1-e^{i\lambda \al}) /\lambda \al $. Also in this case,
as in the Heisenberg-group case, the identification of the element
$\al,\bt\in \mathbb{R}^{2}$ with the element $e^{T(\al,\bt)}$ makes
$\mathbb{R}^{2}$ into a group.

We notice that the elements $e^{-ik\x}e^{iw \x_0}$ which appear in
(\ref{KWeylmap}) are elements of a unitary \kM\ group with $\al=iw$
and $\bt=-ik/\phi(iw)$, in fact  $$
e^{T(\al,\bt)}=e^{T(0,\,\phi(\al)\bt)}e^{T(\al,0)}=e^{-ik\x}e^{iw\x_0}$$
with the following group law for $(w,k')$ \be
(w,k)\cdot(w',k')=(w+w',k+e^{-\lambda w}k') \label{right}\ee This
relation is easily generalized to $1\!+\!3$ dimension \be
(w,k_j)\cdot(w',k_j')=(w+w',k_j+e^{-\lambda w}k_j'),\;\;\;j=1,2,3
\label{right4dim}\ee Thus we have found that the ``plane waves"
$\{e^{-ik_j\x_j}e^{iw\x_0},\;w,k_j\in \mathbb{R}\}$ in
(\ref{KWeylmap}) are the elements of the unitary Lie group for the
\kM\ Lie algebra $\g_\lambda$ with composition law (\ref{right4dim})
and identity given by the element $(0,0)$. We shall refer to this
group as the \kM-group and we shall denote it with the symbol
${\mathcal{U}_\lambda}$.

\section{\label{sec:4}Representation Theory}
In this section we look for irreducible representations of the
\kM\ group ${\mathcal{U}_\lambda}$ on Hilbert spaces.

Let us remind that a representation of a group $G$ on a Hilbert
space ${\mathcal{H}}$ is a map $\rho:\;G\to L({\mathcal{H}})$ of
$G$ into a set of linear operators on ${\mathcal{H}}$ satisfying
the conditions \bea
\rho(ab)&=&\rho(a)\rho(b),\;\;\;a,b\in G\nn\\
\rho(e)&=&\textbf{1}\nn\eea where $\{e\}$ is the identity of $G$.
The dimension of the representation is defined as the dimension of
${\mathcal{H}}$.

First of all we notice that ${\mathcal{U}_\lambda}$ is a solvable
group. For any two elements $x,y\in G$ the set of elements
$q=xyx^{-1}y^{-1}$ form a group $Q$ called commutant of $G$. Set
$Q\equiv Q_0$ and $Q_1$ the commutant of $Q_0$, $Q_2$ the commutant
of $Q_1$ etc... If for some $m$ we have $Q_m=\{e\}$ then the group
$G$ is said solvable. It is easy to check from (\ref{right4dim})
that in the case of ${\mathcal{U}_\lambda}$ the second commutant is
$Q_1=(0,0)$, then ${\mathcal{U}_\lambda}$ is solvable.

We can then apply Lie's theorem which states the following (see~\cite{Barut}):\\
\textbf{Lie Theorem}: \emph{Every finite-dimensional irreducible
representation of a connected topological, solvable group $G$ is
one-dimensional.}

It is easy to see that the only one 1-dimensional irreducible
representation of ${\mathcal{U}_\lambda}$ is realized for
$\lambda=0$ and it coincides with the representation of the Abelian
group on the real line $\mathbb{R}$. However, we are interested in
irreducible representations of \kM\ for $\lambda\neq 0$ then we need
to search for them in infinite dimensional Hilbert spaces.

In the first part of this Section, we show how to obtain some
irreducible representations of ${\mathcal{U}_\lambda}$ using the JS
maps. In particular, we give the explicit expression of two classes
of Schr\"odinger representation of ${\mathcal{U}_\lambda}$ on
$L^2(\mathbb{R})$.
In the second part we use the technique of induced representations
to prove that all the possible irreducible representations of
${\mathcal{U}_\lambda}$ are unitarily equivalent to these two
classes of Schr\"odinger representations.

\subsection{Construction of \kM\ representations through
Jordan-Schwinger maps}

Here we show that the representation of the generators $Q_j,P_j$
of the Heisenberg algebra provides us an easy  way of obtaining
representations of the \kM\ generators $\x_\mu$. Our idea is based
on the introduction of Jordan-Schwinger maps $\pi$ from the
coordinates $\x_\mu$ of \kM\ to the generators $Q_j,P_j$  $$
\x_\mu=\pi^{-1}(Q_j,P_j)$$

These maps can be extended to the exponential functions
$e^{-ik_j\x_j}e^{iw\x_0}$ obtaining the representations of the
\kM\ group on the Hilbert spaces on which the Heisenberg algebra
is represented.

The JS map was originally introduced by Schwinger to deal with
angular momentum in terms of harmonic-oscillator creation and
annihilation operators $(a_j,a^*_j),\;\;j=1,2$, satisfying the
standard commutation relations $[a_j,a^*_k]=\delta_{jk}$. However,
since the creation and annihilation operators ($a,a^*$) can be
connected to the Quantum Mechanics generators ($[Q,P]=i$) by the
following relations
$$Q=\frac{a+a^*}{\sqrt{2}},\;\;\;P=\frac{a-a^*}{i\sqrt{2}}$$
we prefer to write the JS maps in terms of ($Q,P$).

A realization of $SL(2)$ in terms of $Q,P$
 is given by: \bea
 J_1=\frac{1}{2}(Q_1Q_2+P_1P_2 )\;\;\;J_2=\frac{1}{2}(Q_1P_2-Q_2P_1)\;\;\;J_3=\frac{1}{4}(Q_1^2-Q_2^2+P_1^2-P_2^2)\nn
 \eea
we have in fact $$[J_j,J_k]=i\epsilon_{jkl}J_l$$

In Ref.~\cite{selene} a generalization of JS map is given for all
three dimensional Lie-algebras.

The following form of the JS map $\pi_+$ is suggested for the
generators of $1+1$-dimensional \kM\ Lie algebra
\be \pi_+(\x_0)=\lambda^2 P,\;\;\;\;\;\pi_+(\x)=\lambda
e^{-\frac{Q}{\lambda}} \label{JS+} \ee This map is easily
extendable to the $1+3\,$-dimensional case but we prefer to work
in two dimensions and then to extend the result to $1+3$
dimensions.

The JS map above is a Lie-algebra homomorphism \be
[\pi_+(\x_0),\pi_+(\x)]=\lambda^3[P,e^{-\frac{Q}{\lambda}}]=i\lambda^2
e^{-\frac{Q}{\lambda}}=i\lambda\pi_+(\x) \ee
and in particular, it is a Lie-algebra *-homomorphism \be
[\pi_+(\x_\mu)]^*=\pi_+(\x_\mu)=\pi_+(\x_\mu^*) \ee Thus this map
reflects the hermitian property of $\x_\mu$.

This map is not unique. A wide class of JS maps can be constructed
from (\ref{JS+}) by canonical transformations $$ Q\to Q',\; P\to P'
:\;\;\; [Q',P']=i$$ In fact, the canonical condition $[Q',P']=i$
assures that $[\pi'(\x_0),\pi'(\x)]=i\lambda \pi'(\x)$.

However another class of JS maps exists which is not possible to
connect with the previous one. In fact consider the map \be
\pi_-(\x_0)=\lambda^2P,\;\;\;\pi_-(\x)=-\lambda\,
e^{-\frac{Q}{\lambda}}\label{JS-}\ee which also reproduces the
\kM\ commutation relations. Despite the two maps differ by a sign,
it is not possible to find a canonical transformation $Q,P\to
Q',P'$ that maps $\pi_-(\x)$ into $\pi_+(\x)$. As we shall see
later, this fact has implications in the classification of the
irreducible representations of the \kM\ group.\\

The JS maps allow us to represent the operators $\x_\mu$ on the
Hilbert spaces in which the generators $Q,P$ of the Heisenberg
algebra are represented.

Here we focus on the Schr\"odinger representation of the
Heisenberg algebra, without worrying about other representations.
The Stone-Von Neumann theorem states in fact that once we have
chosen a non-zero Planck's constant, the irreducible
representation of the Heisenberg algebra is unique
 (see~\cite{Folland}).
The Hilbert representation space of the Schr\"odinger
representation is $L^2(\mathbb{R})$, the vector space of all
square integrable complex-valued functions on $\mathbb{R}$,
\be L^2(\mathbb{R})=\{\psi :\mathbb{R}\to \mathbb{C},\;\;
||\psi||^2=\int_{\mathbb{R}} dx\,\overline{ \psi(x)}\psi(x)<\infty
\} \ee The generators $Q,P$ are represented on $L^2(\mathbb{R})$ in
the following way (see Appendix~\ref{app:3} for details about our
notation): \bea Q\psi(x)=x\psi(x) &&P\psi(x)=-i\partial_x\psi(x)\nn
\eea

Through the map (\ref{JS+}) we get the Schr\"odinger
representation of $\x_\mu$ on $L^2(\mathbb{R})$
 \bea \x_0\,
\psi(x)&=&-i\lambda^2 \partial_x\,\psi(x)\nn\\
\x\,\psi(x)&=&\lambda e^{-\frac{x}{\lambda}}\,\psi(x)\nn \eea
where $\psi(x)\in L^2(\mathbb{R})$.

If we extend the  map $\pi_+$ as *-homomorphism to the Universal
Enveloping Algebra (UEA) of \kM\ Lie-algebra $\g_\lambda$ such
that
$$ \pi_+(\x_0^m\x^n)=\pi_+(\x_0)^m\pi_+(\x)^n $$ we obtain the
following map $\pi_+$ on the elements of the \kM\ group
${\mathcal{U}_\lambda}$ \be
\pi_+(e^{-ik\x}e^{iw\x_0})=e^{-ik\pi_+(\x)}e^{iw\pi_+(\x_0)}
=e^{-i\lambda ke^{-\frac{Q}{\lambda}}}e^{i\lambda^2 wP} \ee

By using the Quantum Mechanics formalism summarized in Appendix
\ref{app:3}, we can represent the operator $\pi_+$ on $
L^2(\mathbb{R})$ and obtain \bea e^{-i\lambda
ke^{-\frac{Q}{\lambda}}}e^{i\lambda^2 wP}\psi(x)
&=&e^{-i\lambda ke^{-\frac{x}{\lambda}}}\psi(x+\lambda^2w )\nn
\eea In this way we have obtained a map $\rho_+:\;
{\mathcal{U}_\lambda}\to U(L^2(\mathbb{R}))$ from the \kM\ group
${\mathcal{U}_\lambda}$ to unitary operators on $L^2(\mathbb{R})$
\be \rho_+(w,k)\psi(x)=e^{-i\lambda
ke^{-\frac{x}{\lambda}}}\psi(x+\lambda^2 w) \label{rho+}\ee

The same procedure for the map $\pi_-$ (\ref{JS-}) leads to the
following unitary representation of $\mathcal{U}_\lambda$ on
$L^2(\mathbb{R})$ \be \rho_-(w,k)\psi(x)=e^{i\lambda
ke^{-\frac{x}{\lambda}}}\psi(x+\lambda^2w)\label{rho-}\ee

At this level of discussion we have only exhibited two kinds of
representations of $\mathcal{U}_\lambda$ and we have shown how they
can be obtained through JS maps. We do not want to analyze here the
problem of classification of all representations of the \kM\ group.
In the next section, however, we shall prove that any irreducible
representation $\rho$ can be connected by a unitary transformation
$U$ to the representation $\rho_+$ or $\rho_-$ which exhaust the
classification of all irreducible representations of the \kM\ group.

It is easy to prove that these representations are irreducible. We
show it in the case of $\rho_+$. Suppose that it exists a subspace
$V\subset L^2(R)$ left invariant by $\rho_+(w,k)$. Consider a
nonzero vector $\psi\in V$ and a vector $\phi\in V_\bot$ where
$V_\bot$ is orthogonal to $V$. Since $\rho_+(w,k)\psi\in V$, we have
that $$ \langle\phi,\rho_+(w,k)\psi\rangle=0, \;\;\;\forall w,k\in
\mathbb{R} $$ and using the explicit representation of $\rho_+(w,k)$
$$ \int dx\, e^{-i\lambda k
e^{-\frac{x}{\lambda}}}\overline{\phi(x)}\psi(x+\lambda^2w)=0\;\;\;\forall
w,k\in \mathbb{R}$$ which can be rewritten as
$$ -\lambda\int dy \frac{\theta(y)}{y}\, e^{-ik
y}\overline{\phi(\lambda\log(\frac{\lambda}{y}))}\psi(\lambda\log(\frac{\lambda
e^{\lambda w}}{y}))=0\;\;\;\forall w,k\in \mathbb{R}$$ which means
that $$
\frac{\theta(y)}{y}\overline{\phi(\lambda\log(\frac{\lambda}{y}))}\psi(\lambda\log(\frac{\lambda
e^{\lambda w}}{y}))=0\;\;\;\forall w\in \mathbb{R}, \forall y\in
\mathbb{R}_+$$ setting $y=\lambda e^{-\lambda^{-1} x}$ we get $$
\overline{\phi(x)}\psi(x+\lambda^2 w)=0\;\;\;\forall
w\in \mathbb{R}, \forall x\in \mathbb{R}$$

Since $\psi\neq 0$ and $w$ is arbitrary then $\phi=0$ and $V\equiv
L^2(\mathbb{R})$.


\subsection{Construction of \kM\ representations through the
technique of the induced representations}

As we will show below the \kM\ group  (\ref{right}) is a
semidirect-product group and in the case of semidirect products the
technique of induced representations is very powerful. In fact there
is a Theorem~\cite{Barut} stating that every IUR of a semidirect
product $G$ is induced from the representation of a proper subgroup
of $G$. Thus, this technique allows us to classify all IURs
representations of the group.

We say that a group $G$ is the semidirect product of two proper
subgroups $S,N\subset G$, and we denote it as $G=S \ltimes N$,
if\\
i) the subgroup $N$ is normal, i.e. for any $g\in G$ and $n\in N$
the element $gng^{-1}$ is still in $N$. \\
ii) every element $g\in G$ can be written in one and only one way
as $g=ns$, where $n\in N$ and $s\in S$.

The group $S$ acts on $N$ by conjugation. Let $s\rhd n$ denotes
the action of $s\in S$ on $n\in N$: \be s\rhd n=s ns^{-1}\ee

We can write the elements of $G$ as ordered pairs $g=(s,n)$, and
imagine that the subgroups $S=(s,0)$ and $N=(0,n)$. The
composition law of the semidirect product $G=S \ltimes N$ takes
the following form \be (s_1,n_1)(s_2,n_2)=(s_1\cdot s_2,\;n_1\cdot
s_1\rhd n_2) \ee where the symbol $\cdot$ denotes the composition
law in $S$ and $N$.

In the case of the \kM\ group ${\mathcal{U}_\lambda}$ (\ref{right}),
we can identify the groups $S$ and $N$ as the two Abelian subgroups
$S=\{(w,0), \;w\in \mathbb{R}\}$ and $N=\{(0,k),\;k \in
\mathbb{R}\}$. The composition law $\cdot$ is the standard sum. The
action of $S$ on $N$ is given by \be s\vartriangleright
n=sns^{-1}=(w,0)(0,k)(-w,0)=(0,e^{-\lambda w}k )\ee Identifying the
element $(w,0)\equiv w$ and $(0,k)\equiv k$ we write the action
above as \be w\vartriangleright k=e^{-\lambda w}k\ee

The construction of induced representations of a semidirect
product $S\ltimes N$ is based on two notions: \begin{itemize}
\item the \underline{orbits} $\hat{O}_{\hat{n}}$ which are the set
of all points in $\hat{N}$ (the dual space of $N$) connected to a
point $\hat{n}\in \hat{N}$ by the action of $S$ on $\hat{n}$.
\item the \underline{stability group} $S_{\hat{O}_{\hat{n}}}$
which is a subgroup of $S$ which leaves invariant the orbit
$\hat{O}_{\hat{n}}$ ($\hat{k}\vartriangleleft w=\hat{k}$, for any
$w\in S_{\hat{O}_{\hat{n}}}$, $\hat{k}\in \hat{O}_{\hat{n}}$).
\end{itemize}

By definition, the dual space of a group is the set of equivalence
classes of all continuous, irreducible unitary representations of
the group. Since in the \kM\ case $N$ is Abelian, any irreducible
representation of it is one-dimensional, then it is represented in
the complex space $\mathbb{C}$. The unitary irreducible
representation of $N$ consists then of the functions $\hat{n}: N \to
\mathbb{C}$ such
that \bea |\hat{n}(n)|&=&1\nn\\
\hat{n}(n_1n_2)&=&\hat{n}(n_1)\hat{n}(n_2)\nn\eea Notice that this
is just the definition of characters of $N$, thus the space
$\hat{N}$ coincides with all characters of $N$. In our case
$N=\{(0,k)\}\sim \{k\}$ and we denote with $\hat{k}\in \hat{N}$ the
dual of $k\in N$. As $N$ is isomorphic to $\mathbb{\mathbb{R}}$,
every character $\hat{k}(\cdot)$ has the form \be
\hat{k}(k)=e^{i\hat{k}k}, \;\;\hat{k}\in \mathbb{R}\ee and also
$\hat{N}$ is isomorphic with $\mathbb{R}$ (and with $N$ itself).

Using the duality between $N$ and $\hat{N}$ we find the action of
$S$ on $\hat{N}$ \be \hat{k}(w\vartriangleright
k)=e^{i\hat{k}(w\vartriangleright k)}=e^{i\hat{k}e^{-\lambda
w}k}=e^{i(\hat{k}e^{-\lambda w})k}=(\hat{k}\vartriangleleft
w)(k)\ee Thus the action of an element $w\in S$ on $\hat{k}\in
\hat{N}$ is given by \be \hat{k}\vartriangleleft w=e^{-\lambda
w}\hat{k} \ee

The set of all $\hat{k}\vartriangleleft w$ for a given $\hat{k}\in
\hat{N}$ and $\forall w\in S$, is called orbit of the character
$\hat{k}$, and is denoted by $\hat{O}_{\hat{k}}$. Two orbits
$\hat{O}_{\hat{k_1}}$ and $\hat{O}_{\hat{k_2}}$ either coincide or
are disjoint. Thus the dual space $\hat{N}$ decomposes into
nonintersecting sets.

In our case, the orbit $\hat{O}_{\hat{k}}$ of the point $\hat{k}\in
\hat{N}$ is the set $\hat{O}_{\hat{k}}=e^{-\lambda w}\hat{k}$ for
all $w \in \mathbb{R}$. We can identify each orbit with its value
$\hat{k}= l$ at $w=0$. Thus, we can distinguish two types of orbits:
\begin{enumerate} \item the orbits with $ l \neq 0$  \item and the orbits with
$ l =0$
\end{enumerate}
The first set of orbits does not have any stability group
$S_{\hat{O}}=\{0\}\subset S$, while the second orbit is a fixed
point under the action of any element of $S$, then $S_{\hat{O}}=S$.


\bigskip
\bigskip
\bigskip

In order to construct and classify the IUR of \kM\ we use the
following theorem which states that every irreducible
representation of a group $G$ of semidirect-product type is
induced from the
representation of a subgroup $K\subset G$. \\
\textbf{IUR Theorem}. \emph{Let $G$ be a regular\footnote{We say
that $G$ is a regular semidirect product of $N$ and $S$ if $\hat{N}$
contains a countable family $Z_1,Z_2,...$ of Borel subsets, each a
union of $G$ orbits, such that every orbit in $\hat{N}$ is the
intersection of the members of a suitable family
$Z_{n_1},Z_{n_2},...$ containing that orbit. One can show that the
regularity condition is fulfilled by $\mathcal{U}_\lambda$.} ,
semidirect product $S\ltimes N$ of separable, locally compact groups
$S$ and $N$, and let $N$ be Abelian. Let $T$ be an IUR of $G$.
Thus:}

\emph{ i) One can associate with $T$ an orbit $\hat{O}$ in
$\hat{N}$, the dual space of $N$. Each orbit has a stability group
$S_{\hat{O}}$ }

\emph{ ii) The representation $T$ is unitarily equivalent to a
representation $\rho^L$ induced by $L$, the irreducible
representation of the group $S_{\hat{O}}\ltimes N$.}

\emph{iii) The representation $\rho^L$ in ii) is irreducible.}
\\
(for a proof see~\cite{Barut})

According to this Theorem, every IUR of the \kM\ group (\ref{right})
is a representation induced by an IUR of the stability subgroup
$S_{O}\ltimes N$ associated with the orbits $ l \neq0$ or $ l =0$.

In Appendix~\ref{app:1} the explicit form of the representation
$\rho^L_{(w,k)}$ of the element $(w,k)\in \mathcal{U}_\lambda$ is
constructed:

\begin{enumerate}
\item Case $ l \neq 0$. The representation is induced by the
representation $L$ of $\{0\}\ltimes S=S\sim \mathbb{R}$, and takes
the form \be\rho^L_{(w,k)}\psi(x)=e^{ik l  e^{-\frac{
x}{\lambda}}}\psi(x+\lambda^2w) \label{rep1}\ee where $\psi\in
L^2(\mathbb{R}; \mathbb{C})$. This representation is parametrized
by the
parameter $ l \neq 0$.\\

\item Case $ l =0$. The representation is induced by the
representation $L$ of $S\ltimes N=G$ and corresponds to the
one-dimensional representation, that is the character
\be\rho_{(w,k)}^L=e^{iwc}I,\;\;\; c\in \mathbb{R}\ee
\end{enumerate}

We discuss case 1 in more detail.\\
The parameter $l$ has the dimension of a length and we can write it
using the natural length scale of our model \emph{i.e.} the
noncommutativity parameter $\lambda$, and since $l\neq 0$, we write
\be l=\pm e^{\lambda^{-1} \al} \lambda\;\;\;\;\al\in \mathbb{R}\ee
We split the discussion of case 1 into two subcases characterized by
the positive and negative signs of the parameter $l$. Let us focus
on the first subcase (l>0) $$ \rho^{\al}_{(w,k)}\psi(x)=e^{i \lambda
ke^{\lambda^{-1} (\al-x)}}\psi(x+\lambda^2w)
$$
Notice that the case $\al=0$ reproduces the Schr\"odinger
representation $\rho_-(w,k)$ obtained with the JS map (\ref{JS-})
$$
\rho^{\al=0}_{(w,k)}\psi(x)=\rho_-(w,k)\psi(x)
$$

Let us introduce now the unitary transformation \be
U\psi(x)=\psi(x+\al)\ee We see that \bea
U\rho_-(w,k)U^*\psi(x)&=&U\rho_-(w,k)\psi(x-\al)=Ue^{i \lambda ke^{
-\lambda^{-1}(x-\al)}}\psi(x-\al+\lambda^2w)\nn\\
&=&e^{i \lambda ke^{ \lambda^{-1}(\al-x)}}
\psi(x+\lambda^2w)=\rho^{\al}_{(w,k)}\psi(x)\eea This result shows
that any representation $\rho^\al_(w,k)$ is unitarily equivalent to
the Schr\"odinger representation $\rho_-(w,k)$ obtained with the JS
map (\ref{JS-}).

The same thing happens for the second subcase where the
representations with different $\al$ are all unitarily equivalent to
the representation $\rho_+(w,k)$ obtained with the JS map
(\ref{JS+}).

At the end of the day we found that for the case $l\neq0$ (which is
the one we are interested in) the irreducible representations of the
\kM\ group are all unitarily equivalent to the two inequivalent
representations $\rho_+(w,k)$ and $\rho_-(w,k)$ obtained with the JS
maps $\pi_\pm$ \be U^*\rho^{(\al,\pm)}_{(w,k)}U=\rho_\mp (w,k)\ee

\section{\label{sec:5}C$^*$-ALGEBRAS \label{Cstaralgebras}}

\subsection{C$^*$-algebra of Weyl operators }
As we mentioned in the Introduction, the Weyl quantization of a
function $f(q,p)$ of the classical phase space $\mathbb{R}^{2}$ is
the linear map
$$ W(f(q,p))=\int d\al d\bt \tilde{f}(\al,\bt) e^{\al Q+i\bt P} $$
where $\tilde{f}$ is the Fourier transform of $f$, and $\tilde{f}\in
L^1(\mathbb{R}^{2})$. Using the Schr\"odinger representation of
$Q,P$ , $W(f)$ can be represented on $L^2(\mathbb{R})$ as follows
 \bea
 W(f)\psi(x)&=&\int d^n\al d^n\bt \tilde{f}(\al,\bt)e^{i\al x}
e^{i\al\bt/2}\psi(x+\bt)\label{mapW}\eea This can be written by
introducing the kernel $K(x,y)$ of the operator $W(f)$\bea
W(f)\psi(x)&=&\int dy K(x,y)\psi(y) \nn\eea with the kernel given
by $$ K(x,y)=\int d\al\,\tilde{f}(\al,y-x)\,e^{i\al(x+y)/2}$$
Since $\tilde{f}\in L^1(\mathbb{R}^2)$, then the operators $W(f)$
are bounded operators on $L^2(\mathbb{R})$ $$ ||W(f)||\leq
||\tilde{f}||_1 $$

The set all bounded operators $W(f)$ on $L^2(\mathbb{R})$ equipped
with the standard product, the norm $$ ||W||=\sup
\{\frac{||W(f)\psi(x)||_2}{||\psi||_2}: \psi\in L^2(\mathbb{R})\}
$$ and the involution given by the adjoint operation on
$L^2(\mathbb{R})$, is a $C^*$-algebra. We show now that the
$C^*$-algebra generated by $W(f)$ is a $C^*$-algebra of compact
operators on $L^2(\mathbb{R})$.

\textbf{Theorem}: \emph{The operator $W(f)$ defined by
(\ref{mapW}) is a compact operator on $L^2(\mathbb{R})$ for all
$\tilde{f}\in L^1(\mathbb{R}^2)$.}

Proof:\\
Let us show first that if  $\tilde{f}\in L^2(\mathbb{R}^2)$ then
$W(f)$ is the set of Hilbert-Schmidt operators, \emph{i.e.}  $$ tr
[W^*(f)W(f)]<\infty $$

The trace can be computed with the kernel in the following way
\bea tr [W^*(f)W(f)]&=&\int dx dy\, K_{W(f)}(x,y)\,
K_{W^*(f)}(y,x)\nn\\
&=&\int dx dy\, K_{W(f)}(x,y)\, \overline{K_{W(f)}(x,y)}=\int dx
dy\, |K_{W(f)}(x,y)|^2\nn \eea

Since the map $f\to K_f$ is unitary, hence norm preserving, we
have that \bea \int dx dy\,|K_{W(f)}(x,y)|^2=\int d\al d\bt
|\tilde{f}(\al,\bt)|^2<\infty  &\forall \tilde{f}\in
L^2(\mathbb{R}^2)\nn\eea

Thus, $W(f)$ is compact for $f\in L^1\cap L^2$, hence for $f\in
L^1$ since $||W(f)||\leq ||f||_1$ and the norm limit of compact
operators is compact~\cite{Folland}.

\subsection{$C^*$-algebra of \kM\ operators}
The linear map (\ref{KWeylmap}) \be \mathcal{W}(f)=\int d^2k\;
\tilde{f}(k)\,
e^{-ik\x}e^{iw\x_0} \ee 
can be represented on $L^2(\mathbb{R})$ using the representation
(\ref{rep1}) (setting $ l =\lambda$ to be simple) \bea
\mathcal{W}(f)\psi(x)&=&\int d^2k\; \tilde{f}(k)\,
e^{i\lambda ke^{-\frac{x}{\lambda}}}\psi(x+\lambda^2w)\label{kmapW}
 \eea Since $\tilde{f}\in
L^1(\mathbb{R}^2)$ the operators $\mathcal{W}(f)$ are bounded
operators on $L^2(\mathbb{R})$ \be ||\mathcal{W}(f)||\leq
||\tilde{f}||_1 \ee

The set all bounded operators $\mathcal{W}(f)$ on
$L^2(\mathbb{R})$ equipped with the standard product, the norm \be
||\mathcal{W}||=\sup \{\frac{||\mathcal{W}\psi(x)||}{||\psi||}:
\psi\in L^2(\mathbb{R})\} \ee and the involution given by the
adjoint operation on $L^2(\mathbb{R})$, is a $C^*$-algebra.

Now we ask under which hypotheses the \kM\ operators
$\mathcal{W}(f)$ are compact operators. Following the case of Weyl
operators $W(f)$ we compute the trace \be \tr[
\mathcal{W}^*(f)\mathcal{W}^*(f)]=\int dx dy
|K_{\mathcal{W}(f)}(x,y)|^2\ee where the kernel
$K_{\mathcal{W}(f)}$ is given by \be K_{\mathcal{W}(f)}(x,y)=\int
dwdk\; \tilde{f}(w,k)\,
e^{i\lambda ke^{-\frac{x}{\lambda}}}\delta(x-y+\lambda^2w)\ee

A direct calculation shows that \bea \int dx dy
|K_{\mathcal{W}(f)}(x,y)|^2 &=&\int dxdy \,dwdk dw'dk'\;
\tilde{f}(w,k)\overline{\tilde{f}(w',k')}
e^{i\lambda ke^{-\frac{x}{\lambda}}}e^{-i\lambda k'e^{-\frac{x}{\lambda}}}\nn\\
&&\delta(x-y+\lambda^2w)\delta(x-y+\lambda^2w')\nn \\
&=&\lambda^{-2}\int dxdy \,dwdk dw'dk' dz dz'\;
\tilde{f}(w,k)\overline{\tilde{f}(w',k')}
e^{i\lambda ke^{-\frac{x}{\lambda}}}e^{-i\lambda k'e^{-\frac{x}{\lambda}}}\nn\\
&&e^{-iz\frac{x-y}{\lambda^2}}e^{iz'\frac{x-y}{\lambda^2}}e^{-izw}e^{izw'}\nn\\
&=&\lambda^{-2}\int dxdy \, dz dz'\;
f(z,e^{-\frac{x}{\lambda}})\overline{f(z',e^{-\frac{x}{\lambda}})}e^{-i(z-z')\frac{x}{\lambda^2}}e^{i(z-z')\frac{y}{\lambda^2}}\nn
\\
&=&\int dx \, dz dz'\;
f(z,e^{-\frac{x}{\lambda}})\overline{f(z',e^{-\frac{x}{\lambda}})}e^{-i(z-z')\frac{x}{\lambda^2}}\delta(z-z')\nn\\
&=&\int dz \, dx \; |f(z,e^{-\frac{x}{\lambda}})|^2=\int dz
dx\frac{\lambda\theta(x)}{x} \; |f(z,x)|^2\nn \eea Thus the \kM\
operators $\mathcal{W}(f)$ are Hilbert-Schmidt operators if $f\in
L^2(\mathbb{R}^2, d\mu)$ with $d\mu=\lambda dz dx \theta(x)/x$.
And, as we showed for Weyl operators, $\mathcal{W}(f)$ is compact
for $f\in L^1\cap L^2$, hence for $f\in L^1(\mathbb{R}^2, d\mu)$
since $||\mathcal{W}(f)||\leq ||f||_1$ and the norm limit of
compact operators is compact.

\section{\label{sec:6}Construction of a cyclic integration \label{cyclic}}
In this section we focus on one particular choice of representation
of $\x_\mu$ among the ones illustrated above \bea
\pi(\x_0)\psi(x)&=&-\frac{\lambda}{2}\sum_{j=1}^3[Q_jP_j+P_jQ_j]\psi=i\lambda\sum_{j=1}^3(x_j\partial_j+\frac{1}{2})\psi\nn\\
\pi(\x_j)\psi(x)&=&Q_j\psi(x)=x_j\psi(x)\label{standard}\eea with
the factor $\lambda$ for the correct physical dimensions.

The JS map (\ref{standard}) is connected with the JS maps
$\pi_\pm$ through the transformation \be Q\to Q'=\pm
\lambda e^{-\frac{Q}{\lambda}}\;\;\;\;\;\;P\to P'=\mp(\lambda
P-\frac{i}{2})e^{\frac{Q}{\lambda}}\ee We notice that the
transformation $Q\to e^{-\frac{Q}{\lambda}}$ maps the Hermitian
operator $Q$ into a positive operator. The other transformation
$Q\to -e^{-\frac{Q}{\lambda}}$ instead maps $Q$ into a negative
operator. Thus the JS map $\pi$ is connected to $\pi_+$ for the
spectrum $x>0$ and with $\pi_-$ for the
spectrum $x<0$.\\

We show in Appendix~\ref{app:2} that the two representations $\rho$
and $\rho_+$ of the \kM\ group, obtained from the JS maps $\pi$ and
$\pi_+$, are unitarily equivalent, i.e., there is a unitary
transformation $U:\;L^2(\mathbb{R})\to L^2(\mathbb{R}_+)$, such that
\be \rho(w,k)=U\rho_+(w,k)U^*\ee

Now we use the representation in Eq. (\ref{standard}) of the \kM\
algebra in order to obtain a rule of integration in \kM\
noncommutative spacetime.

Let us remind that the trace of the operators $\mathcal{W}(f)$ is
defined as \bea
 \tr{\mathcal{W}(f)}&=&\int dx K(x,x)\label{trace}\eea
 where $K(x,x)$ is the kernel of $\mathcal{W}(f)$
 $$ \mathcal{W}(f)\psi(x)=\int dy K(x,y)\psi(y)$$ It easy to
see that for the representation (\ref{standard}) \bea K(x,x)=\int
dw dk \tilde{f}(w,k)e^{-\frac{\lambda
w}{2}}e^{-ikx}\delta(e^{-\lambda w}x-x)=\int dw dk
\frac{1}{\lambda|x|}\tilde{f}(w,k)e^{-\frac{\lambda
w}{2}}e^{-ikx}\delta(w)\nn\eea and thus the trace is \bea
\tr{\mathcal{W}(f)}=\int dx dk
\frac{1}{\lambda|x|}\tilde{f}(0,k)e^{-ikx}\nn\eea  This result can
be easily extended to $3+1$ dimensions \bea
\tr{\mathcal{W}(f)}&=&\int \frac{d^3x}{\lambda^3 |\vec{x}|^3} d^3k\;
\tilde{f}(0,\vec{k})e^{-i\vec{k}\vec{x}}\nn\eea and, using the
Fourier transform formulas, we obtain the final result: \be
\tr{\mathcal{W}(f)}=\int
\frac{dx_0d^3x}{\lambda^3|\vec{x}|^3}f(x_0,\vec{x})\ee In this way we can
see that the trace (\ref{trace}) of the \kM\ operators
$\mathcal{W}(f)$ corresponds to the cyclic integral on \kM\
spacetime obtained in \cite{Agostini:2004cu}
 and \cite{Dimitrijevic:2003wv}. In these papers, where the star-product ($*$) formalism was used,
 a cyclic integral was constructed by introducing a non-trivial
 measure
 \be
 I[\mathcal{W}(f)\mathcal{W}(g)]\equiv\int d^4x \,\mu(x)\,(
 f(x)*g(x))\ee
and, in particular, in \cite{Agostini:2004cu} the explicit
expression  $\mu(x)=\frac{1}{|\vec{x}|^3}$ was obtained by
imposing
the cyclicity condition  \be 
\int d^4x\,\mu(x)\,[f*g-g*f]=0\ee Of course, this condition is
naturally fulfilled by the request that the integral in \kM\
correspond to a trace of operators on Hilbert space \be
\tr[\mathcal{W}(f)\mathcal{W}(g)]=\tr[\mathcal{W}(g)\mathcal{W}(f)]
\ee

\section{\label{sec:7}Summary and Outlook}
The Weyl formulation of Quantum Mechanics is based on the
introduction of (Weyl) operators. Looking for a formulation of a
field theory in \kM\ noncommutative spacetime a generalization of
Weyl operators has been introduced. In this paper we studied the
problem of the representations of these $\kappa$-generalized Weyl
operators on Hilbert spaces. They are essentially Fourier expansions
in terms of elements of the Lie group of \kM. Therefore it has been
sufficient to study the representations of the \kM\ group.
Introducing Jordan-Schwinger maps between Quantum Mechanics
generators and \kM\ coordinates the representations of \kM\ group is
immediately obtained. However many JS maps, related by canonical
transformations, are possible giving rise to several representations
of the \kM\ group. The problem of the classification of all unitary
irreducible representations has been solved by the technique of
induced representations, and we found that all IURs of the \kM\
group are essentially led back to two types of Schr\"odinger
representations on $L^2(R)$. As a natural application of the
representation theory one can construct an integration on \kM\
spacetime as trace of operators. This integration is automatically
cyclic in the arguments and reproduces the result obtained in
literature using a different approach. The cyclic integration
involves the presence of a nontrivial measure $d\mu$, which seems a
characterization of \kM\ with respect to noncommutative spaces with
canonical structure (including the Quantum Mechanics phase space as
well). The approach of the present paper, based on the
representations of \kM\ fields as operators on Hilbert space, opens
a new way to study a field theory on \kM\ noncommutative spacetime.
As one interesting application of this approach one could construct
the area and volume operators in \kM\, and compute their spectra.
This could provide a comparison with the results of Loop Quantum
Gravity where a detailed computation of the area/volume spectra has
been done.

\section*{ACKNOWLEDGMENTS} I am grateful for fruitful conversations
with G.~Amelino-Camelia, E.~Blanchard, S.~Doplicher, D.E~Evans and
F.~Lizzi. I would also like to thank L.~Jonke, L.~M\"{o}ller and
E.~Tsouchnika for helpful interactions and proofreading of this
article. My work is supported by ``Quantum Spaces-Noncommutative
Geometry" EU Network.



\appendix

\section{\label{app:1}Construction of induced representation}

In this appendix we construct explicitly the induced representations
of \kM\ group.

Let $L_h$ be a unitary representation of a closed subgroup
$H\subseteq G$ in a separable Hilbert space ${\mathcal{H}}$. Let
$d\mu(x)$ a quasi-invariant measure\footnote{A quasi-invariant
measure $d\mu(x)$ on $X=G\setminus H$ is a positive measure such
that $d\mu(xg)=\rho(x)d\mu(x)$ for any $g\in G$. The function
$\rho(x)$ is said Radom-Nikodym derivative. } on the homogeneous
space $X=H\diagdown G=\{Hg,\;g\in G\}$ of the right $H$-coset.
Consider the set ${\mathcal{H}}^L$ of all function $u:\;G\to {\mathcal{H}}$ such that:\\
1. $\langle u(g),v\rangle$ is measurable for all $v\in {\mathcal{H}}$.\\
2. $u(hg)=L_hu(g)$, for all $h\in H$ and all $g\in G$.\\
3. $\int_X |\!|u(g)|\!|_{\mathcal{H}}^2 \, d\mu(x)<\infty$.

The map $g_0\to \rho_{g_0}^L$ given by \be \rho_{g_0}^L
u(g)=\sqrt{\frac{d\mu(xg_0)}{d\mu(x)}}\, u(gg_0) \label{indrep}\ee
defines a unitary representation of $G$ in ${\mathcal{H}}^L$.\\
Proof: It easy to see that the map is a representation of $G$ in
${\mathcal{H}}^L$ \bea \rho_{g_1}^L\rho_{g_0}^L
u(g)&=&\sqrt{\frac{d\mu(xg_0)}{d\mu(x)}}\rho_{g_1}^Lu(gg_0)=
\sqrt{\frac{d\mu(xg_0g_1)}{d\mu(x)}}u(gg_0g_1)=\rho_{g_0g_1}^L
u(g) \nn\eea The unitarity is simply proved by noticing that \bea
\int_X|\!|\rho_{g_0}^L
u(g)|\!|^2d\mu(x)=\int_X|\!|u(gg_0)|\!|^2d\mu(xg_0)=\int_X|\!|u(g)|\!|^2d\mu(x)\nn\eea

If we consider $H=S_{\hat{O}}\ltimes N$, where $S_{\hat{O}}$ is
the stability group of the orbit $\hat{O}$, the formula
(\ref{indrep}) can be rewritten making use of the following
Lemmas.

\emph{Lemma 1.} Every IUR $L$ of $S_{\hat{O}}\ltimes N$ is
determined and determines an IUR $L_{S_{\hat{O}}}$ of $S_{\hat{O}}$
$$ L_{(s_{\hat{O}},n)}=\langle n,\hat{n}_f\rangle
L_{s_{\hat{O}}},\;\;\; $$ where $(s_{\hat{O}},n)\in
H=S_{\hat{O}}\ltimes N $, and $ \hat{n}_f$ is the fixed point under
the action of $S_{\hat{O}}$ ($\hat{n}_f\vartriangleleft s_{\hat{O}}=
\hat{n}_f $ for any
$s_{\hat{O}}\in S_{\hat{O}}$.\\
Proof: it is easy to see that \bea
L_{(s_{\hat{O}},n)(s_{\hat{O}}',n')}&=&
L_{(s_{\hat{O}}s_{\hat{O}}',n+s_{\hat{O}}'\vartriangleright n')}=\langle n+s_{\hat{O}}'\vartriangleright n', \hat{n}_f \rangle L_{s_{\hat{O}}s_{\hat{O}}'}\nn\\
&=&\langle n, \hat{n}_f\rangle\langle
s_{\hat{O}}'\vartriangleright n', \hat{n}_f \rangle
L_{s_{\hat{O}}}L_{s_{\hat{O}}'}=\langle n,\hat{n}_f \rangle
L_{s_{\hat{O}}}\langle n',\hat{n}_f \rangle
L_{s_{\hat{O}}'}\nn\\
&=&L_{(s_{\hat{O}},n)}L_{(s_{\hat{O}}',n')}\nn\eea and $$
L_{(s_{\hat{O}},n)}^*=\overline{\langle n,\hat{n}_f
\rangle}L_{s_{\hat{O}}}^{-1}=\langle n^{-1}, \hat{n}
\rangle=L^{-1}_{(s_{\hat{O}},n)}$$

\emph{Lemma 2.} The set of function $u\in {\mathcal{H}}^L$, such
that $u(hg)=L_h u(g)$ for any $h\in H=S_{\hat{O}}\ltimes N$, can
be written as \be u(g)=\langle n, \hat{n} \rangle
u(s),\;\;\;g=(s,n)\label{HL}\ee with the condition
$u(s_{\hat{O}}s)=L_{s_{\hat{O}}}u(s)$ for any $s_{\hat{O}}\in
S_{\hat{O}}$.
\\
Proof: Let $h=(s_{\hat{O}}',n')$ then a simple calculation shows
that \bea
u(hg)&=&u[(s_{\hat{O}}',n')(s,n)]=u[s_{\hat{O}}'s,\;n'+s_{\hat{O}}'\vartriangleright
 n]\nn\\
&=&\langle n'+s_{\hat{O}}'\vartriangleright n, \hat{n}_f \rangle
u(s_{\hat{O}}s)=\langle n',\hat{n}_f\rangle \langle
s_{\hat{O}}'\vartriangleright
n, \hat{n}_f \rangle u(s_{\hat{O}}'s)\nn\\
&=&\langle k, \hat{n}_f \rangle \langle n, \hat{n}_f
\rangle L_{s_{\hat{O}}'}u(s)=\langle k, \hat{n}_f \rangle L_{s_{\hat{O}}'}u(g)=L_{(s_{\hat{O}}',n')}u(g)\nn\\
&=&L_ku(g)\nn\eea

We notice also that the homogeneous space $X=[S_{\hat{O}}\ltimes
N]\diagdown G$ is isomorphic to the space of the orbits $\hat{O}$
$$ X=[S_{\hat{O}}\ltimes N]\diagdown G=[S_{\hat{O}}\ltimes N]\diagdown[S\ltimes
N]\sim S_{\hat{O}}\diagdown S\sim \hat{O} $$ thus \be
\frac{d\mu(xg_0)}{ d\mu(x)}=\frac{d\mu(ss_0)}{d\mu(s)}
\label{measure}\ee

Finally substituting (\ref{HL}) and (\ref{measure}) into
(\ref{indrep}) we obtain for $G\ni g_0=(s_0,n_0)$  \bea
\rho^L_{g_0}u(s)&=&\langle
n,\hat{n}\rangle^{-1}\sqrt{d\mu(ss_0)/d\mu(s)}\langle
n+s\vartriangleright
n_0,\hat{n}\rangle u(ss_0)\nn\\
&=&\sqrt{d\mu(ss_0)/d\mu(s)}\langle s\vartriangleright
n_0,\hat{n}\rangle u(ss_0)\nn\\
&=&\sqrt{d\mu(ss_0)/d\mu(s)}\langle n_0,\hat{n}\vartriangleleft
s\rangle u(ss_0)\label{finalappendix} \eea where
${\mathcal{H}}^L\ni u:\;S\to {\mathcal{H}}$ and satisfy
$u(s_{\hat{O}}'s)=L_{s_{\hat{O}}'}u(s)$, and $\hat{n}\in \hat{O}$
thus it is left fixed under the action of $S_{\hat{O}}$.

This formula gives rise to two representation according to the
values of $\hat{n}$ and the correspondent stability group
$S_{\hat{O}}$.

\begin{enumerate}
\item Case $ \hat{n}=l \neq 0$. The representation is induced by
the representation $L$ of
$\{0\}\ltimes S=S\sim \mathbb{R}$.\\
The character $\langle n_0,\hat{n}\vartriangleleft
s\rangle=e^{in_0le^{-\lambda s}}$. The measure $d\mu(s)$
quasi-invariant on $S$ can be chosen as $ds$ in fact $$
d(ss_0)=d(s+s_0)=ds $$  In this case we have that the factor
$\sqrt{\frac{d\mu(ss_0)}{d\mu(s)}}=1$. The (\ref{finalappendix})
takes the form $$ \rho^L_{(s_0,n_0)}u(s)=e^{in_0 l e^{-\lambda
s}}u(s+s_0) $$ where $u\in L^2(\mathbb{R},\mu; \mathbb{C})$. This
representation is parametrized by the
parameter $l\neq 0$.\\

\item Case $ \hat{n}= 0$. The representation induced by the representation $L$ of $S\ltimes N=G$ \\
In this case the (\ref{finalappendix}) is $$
\rho^L_{(s_0,s_0)}u(s)=u(s+s_0) $$ and since the stability group
is $S$, we have that $u(s+s_0)=L_{s_0} u(s)$ for any $s_0\in S$.
Thus the induced representation is the one-dimensional
representation, that is the character $$
\rho_{g_0}^L=e^{is_0c}I,\;\;\; c\in \mathbb{R}$$

\end{enumerate}

\section{\label{app:2}Example of unitary equivalence between two \kM\ group
representations}

Here we want to show explicitly that the two representations of
the \kM\ group $\rho$ and $\rho_+$, obtained from the JS maps
$\pi$ and $\pi_+$, are unitarily equivalent, i.e., there is a
unitary transformation $U:\;L^2(\mathbb{R})\to L^2(\mathbb{R}_+)$,
such that \be \rho(w,k)=U\rho_+(w,k)U^*\ee

In our case the unitary transformation $U$ is given by \be
\psi(x)\to
U\psi(x)=\frac{\theta(x)}{\sqrt{\frac{x}{\lambda}}}\psi(\lambda
\log\frac{\lambda}{x}) \ee This transformation can be easily proved
to be unitary \bea ||U\psi||^2&=&\int dx |U\psi|^2=\int_0^{\infty}
dx
\frac{\lambda}{x}|\;\psi(\lambda\log\frac{\lambda}{x})|^2=\int_0^{\infty}
d(\lambda\log\frac{\lambda}{x})\,|\psi(\lambda\log\frac{\lambda}{x})|^2\nn\\
&=&\int_{-\infty}^{+\infty} dz |\psi(z)|^2=||\psi||^2\nn\eea

Using the explicit form of the representation $\rho_+$ \bea
\rho_+(w,k)U\psi(x)&=&\frac{\theta(x)}{\sqrt{\frac{x}{\lambda}}}\rho_+(w,k)
\psi(\lambda\log\frac{\lambda}{x})=\frac{\theta(x)}{\sqrt{\frac{x}{\lambda}}}e^{-ikx}
\psi(\lambda[\log\frac{\lambda}{x}+\lambda w])\nn\\
&=&e^{-\lambda w/2}\frac{\theta(e^{-\lambda
w}x)}{\sqrt{\frac{e^{-\lambda w}x}{\lambda}}}e^{-ikx}
\psi(\lambda[\log\frac{\lambda}{x}+\log e^{\lambda w}])\nn\\
&=&e^{-\lambda w/2}\frac{\theta(e^{-\lambda
w}x)}{\sqrt{\frac{e^{-\lambda w}x}{\lambda}}}e^{-ikx}
\psi(\lambda\log\frac{\lambda}{xe^{-\lambda
w}})\nn\\
&=&e^{-\lambda w/2}e^{-ikx}U\psi(e^{-\lambda w}x)\nn\eea thus: \be
\rho(w,k)\psi(x)=e^{-\lambda w/2}e^{-ikx}\psi(e^{-\lambda w}x)\ee

The unitary transformation
$U\psi(x)=\frac{\theta(-x)}{\sqrt{-\frac{x}{\lambda}}}\psi(\lambda
\log\frac{\lambda}{-x})$ allows us to map $\rho$ with the other
representation $\rho_-$ in the negative real line $\mathbb{R}_-$
\bea
\rho_-(w,k)U\psi(x)&=&\frac{\theta(-x)}{\sqrt{\frac{|x|}{\lambda}}}\rho_-(w,k)
\psi(\log\frac{\lambda}{|x|})=\frac{\theta(-x)}{\sqrt{\frac{|x|}{\lambda}}}e^{-ikx}
\psi(\lambda[\log\frac{\lambda}{|x|}+\lambda w])\nn\\
&=&e^{-\lambda w/2}\frac{\theta(-e^{-\lambda
w}x)}{\sqrt{\frac{e^{-\lambda w}|x|}{\lambda}}}e^{-ikx}
\psi(\lambda[\log\frac{\lambda}{|x|}+\log e^{\lambda w}])\nn\\
&=&e^{-\lambda w/2}\frac{\theta(-e^{-\lambda
w}x)}{\sqrt{\frac{e^{-\lambda w}|x|}{\lambda}}}e^{-ikx}
\psi(\lambda\log\frac{\lambda}{|x|e^{-\lambda
w}})\nn\\
&=&e^{-\lambda w/2}e^{-ikx}U\psi(e^{-\lambda w}x)\nn\eea

\section{\label{app:3}Quantum Mechanics formalism} We can see the formalism used
in terms of bra-ket notation (see Sakurai~\cite{sakurai}). We
suppose the element $\psi(x)$ of the Hilbert state be \be
\psi(x)=<x|\psi> \ee

The action of the position-momentum operators over $\psi(x)$ is \be
Q\psi(x)=x\psi(x),\;\;\;P\psi(x)=-i\partial_x\psi(x) \ee

Let us check the commutation relation \bea
QP\psi(x)=<x|QP|\psi>=x<x|P|\psi>=-ix\partial_x\psi(x)\nn\\
PQ\psi(x)=<x|PQ|\psi>=-i\partial_x<x|Q|\psi>=-i\partial_x(x\psi(x))=-i\psi(x)-ix\partial_x\psi(x)\nn
\eea Thus \be [Q,P]\psi(x)=i\psi(x) \ee

Finally, let us compute the action of the operator $e^{iaQ}e^{ibP}$
on the element $\psi(x)$  \bea
e^{iaQ}e^{ibP}\psi(x)&=&<x| e^{iaQ}e^{ibP}|\psi>=e^{iax}<x|e^{ibP}|\psi>\nn\\
&=&e^{iax} e^{b\partial_x}<x|\psi>= e^{iax} \psi(x+b)\nn\\
\eea This formula has been used for getting Equations (\ref{rho+},
\ref{rho-}).



\begin{thebibliography}{99}

\bibitem{DFR} S.~Doplicher, K.~Fredenhagen and J.~E.~Roberts,
Commun.\ Math.\ Phys.\  {\bf 172}, 187 (1995);
Phys.\ Lett.\ B {\bf 331}, 39 (1994).


\bibitem{garay} L.~J.~Garay,
Int.\ J.\ Mod.\ Phys.\ A {\bf 10}, 145 (1995).

\bibitem{Snyder1}
W.~Pauli, in \emph{Letter of Heisenberg to Peierls (1930)},
Scientific Correspondence, Vol II, edited by Karl von Meyenn (
Springer-Verlag, 1985) p.15.

\bibitem{Snyder2}H.~S.~Snyder, \emph{Quantized Space-Time}, Phys.~Rev.~{\bf 71},
38 (1947).

\bibitem{Jackiw:2002tw}
R.~Jackiw, \emph{Lochlainn O'Raifeartaigh, fluids, and noncommuting
fields}, arXiv:physics/0209108.

\bibitem{SW} N.~Seiberg and E.~Witten, 
JHEP {\bf 9909}, 032 (1999).

\bibitem{douglasnovikov} M.R.~Douglas and N.A.~Nekrasov,
Rev.~Mod.~Phys.~{\bf 73}, 977 (2001).

\bibitem{Jackiw:2001dj}
R.~Jackiw,
Nucl.\ Phys.\ Proc.\ Suppl.\  {\bf 108}, 30 (2002).

\bibitem{CDS}
A.~Connes, M.~R.~Douglas and A.~Schwarz, 
JHEP {\bf 9802}, 003 (1998).

\bibitem{MajidRuegg}
S.~Majid and H.~Ruegg, 
Phys.\ Lett.\ B {\bf 334}, 348 (1994).

\bibitem{Maj88MajPhDmaj00} S.~Majid,
Class. Quantum Grav. \textbf{5}, 1587 (1988);
J. Math. Phys. {\bf 41}, 3892 (2000).

\bibitem{aad03}
A.~Agostini, G.~Amelino-Camelia and F.~D'Andrea,
Int. J. Mod. Phys. A {\bf 19}, 5187 (2004).

\bibitem{Dimitrijevic:2003pn}
M.~Dimitrijevic, F.~Meyer, L.~Moller and J.~Wess,
Eur.\ Phys.\ J.\ C {\bf 36}, 117 (2004).

\bibitem{Ballesteros:2003ag}
  A.~Ballesteros, N.~R.~Bruno and F.~J.~Herranz,
  Phys.\ Lett.\  B {\bf 574}, 276 (2003).

\bibitem{lukieAnnPhys} J.~Lukierski, A.~Nowicki and H.~Ruegg,
Ann.~Phys. (NY)~{\bf 243}, 90 (1995).

\bibitem{AmelinoLukNowik}
G.~Amelino-Camelia, J.~Lukierski and A.~Nowicki,
Phys.\ Atom.\ Nucl.\  {\bf 61},
1811 (1998).
Int.\ J.\ Mod.\ Phys.\ A {\bf 14}, 4575 (1999).

\bibitem{dsr1} G.~Amelino-Camelia,
 Nature
{\bf 418}, 34 (2002).

\bibitem{Agostini:2004cu}
A.~Agostini, G.~Amelino-Camelia, M.~Arzano and F.~D'Andrea,
Int.\ J.\ Mod.\ Phys.\ A, {\bf 21}, 3133 (2006).

\bibitem{Agostini:2003dc}
A.~Agostini, Ph. D. thesis, Naples University, 2003,
arXiv:hep-th/0312305.

\bibitem{Madore:2000en} J.~Madore, S.~Schraml, P.~Schupp and J.~Wess,
Eur.\ Phys.\ J.\ C {\bf 16}, 161 (2000).

\bibitem{Minwalla} S. Minwalla, M.V. Raamdonk and N. Seiberg,
JEPH, {\bf 02}, 020 (2000).

\bibitem{selene} J.~M.~Gracia-Bondia, F.~Lizzi, G.~Marmo and P.~Vitale,
JHEP {\bf 0204}, 026 (2002).

\bibitem{AgarwalWolf}
G.~S.~Agarwal and E.~Wolf,
Phys.\ Rev.\ D {\bf 2}, 2161 (1970).

\bibitem{Folland} G. Folland, in \emph{Harmonic Analysis in Phase Space} (Princeton
University Press, 1989).

\bibitem{Agostini:2002de}
A.~Agostini, F.~Lizzi and A.~Zampini,
Mod.\ Phys.\ Lett.\ A {\bf 17}, 2105 (2002).

\bibitem{Kosinski:1999dw}
P.~Kosinski, J.~Lukierski and P.~Maslanka,
Czech.\ J.\ Phys.\  {\bf 50}, 1283 (2000).

\bibitem{Barut} A.O. Barut and P.Ronczka,
in \emph{Theory of group representations and its applications} (Mir,
Moscow, 1980).

\bibitem{Dimitrijevic:2003wv}
M.~Dimitrijevic, L.~Jonke, L.~Moller, E.~Tsouchnika, J.~Wess and
M.~Wohlgenannt,
Eur.\ Phys.\ J.\ C {\bf 31}, 129 (2003).

\bibitem{sakurai}
J.~J.~Sakurai, in \emph{Modern Quantum Mechanics} (Addison-Wesley,
New York, 1994).


\end{thebibliography}
\end{document}